\documentclass[%
reprint,
superscriptaddress,
aps,
pra,
onecolumn
]{revtex4-2}

\usepackage[pdftex]{graphicx, xcolor}
\usepackage{float}
\usepackage{dcolumn}
\usepackage{amsmath,amssymb,bm}
\usepackage{braket}
\usepackage{url}
\usepackage{here}
\usepackage{ulem}

\newcommand{\nm}{\nonumber}

\begin{document}


\title{
Model-based framework for automated quantification of error sources in quantum state tomography
}

\author{Junpei Oba}
\email[Contact author: ]{junpei-oba@mosk.tytlabs.co.jp}
\affiliation{Toyota Central R\&D Labs., Inc., 41-1 Yokomichi, Nagakute, Aichi 480-1192, Japan}

\author{Hsin-Pin Lo}
\affiliation{NTT Basic Research Laboratories, NTT, Inc., 3-1 Morinosato Wakamiya, Atsugi, Kanagawa 243-0198, Japan}
\author{Yasuhiro Yamada}
\affiliation{NTT Basic Research Laboratories, NTT, Inc., 3-1 Morinosato Wakamiya, Atsugi, Kanagawa 243-0198, Japan}

\author{Takayuki Matsui}
\affiliation{Toyota Central R\&D Labs., Inc., 41-1 Yokomichi, Nagakute, Aichi 480-1192, Japan}

\author{Takuya Ikuta}
\affiliation{NTT Basic Research Laboratories, NTT, Inc., 3-1 Morinosato Wakamiya, Atsugi, Kanagawa 243-0198, Japan}
\author{Yuya Yonezu}
\affiliation{NTT Basic Research Laboratories, NTT, Inc., 3-1 Morinosato Wakamiya, Atsugi, Kanagawa 243-0198, Japan}
\author{Toshimori Honjo}
\affiliation{NTT Basic Research Laboratories, NTT, Inc., 3-1 Morinosato Wakamiya, Atsugi, Kanagawa 243-0198, Japan}

\author{Seiji Kajita}
\affiliation{Toyota Central R\&D Labs., Inc., 41-1 Yokomichi, Nagakute, Aichi 480-1192, Japan}
\author{Hiroki Takesue}
\affiliation{NTT Basic Research Laboratories, NTT, Inc., 3-1 Morinosato Wakamiya, Atsugi, Kanagawa 243-0198, Japan}

\date{\today}

\begin{abstract}
High-quality quantum state generation is essential for advanced quantum information processing, including quantum communication, quantum sensing, and quantum computing.
In practice, various error sources degrade the quality of quantum states, and quantum state tomography (QST) is a standard diagnostic tool.
However, in QST, multiple error sources gather in a single density matrix, making it difficult to identify individual error sources.
To address this problem, we propose an automated method for quantifying error sources by combining simulation and parameter optimization to reproduce the experimental density matrix.
We focus on the experimental generation of time-bin entangled photon pairs, for which we model the relevant error sources and simulate the density matrix with adjustable model parameters, thereby optimizing the parameters and minimizing the trace distance to the experimental data.
Optimization of the parameters reduced the trace distance from 0.177 to 0.024, indicating that our modeled error sources explain 86\% of the errors.
Reducing the predicted error sources improves the state quality, consistent with our predictions and thus validating the proposed method.
In addition, the modular structure of this framework makes it applicable to other quantum platforms, such as superconducting qubits, atoms, and solid-state spins.
\end{abstract}

\maketitle

\section{Introduction}
\label{sec:Introduction}

Remarkable progress in quantum technologies has paved the way for practical developments in various quantum information processing applications, such as quantum communication~\cite{Gisin2007, Sidhu2021, Luo2023}, quantum sensing~\cite{Degen2017, Pirandola2018, Aslam2023}, and quantum computing~\cite{O'brien2007, Nielsen2010}.
Fully exploiting these capabilities requires quantum systems of much higher quality than those currently available, and ongoing refinements in superconducting circuits exemplify the steady march toward fault-tolerant thresholds~\cite{Nakamura1999, Yu2002, Barends2014, Google2024}.
However, even minute residual errors can accumulate and compromise deep quantum algorithms~\cite{Gidney2021}.
Higher-quality quantum states not only enhance the reliability and efficiency of quantum operations but also enable more complex applications that surpass the capabilities of current implementations, such as quantum networks~\cite{Kimble2008, Wehner2018}, distributed quantum sensing~\cite{Zhang2021}, and distributed quantum computing~\cite{Cirac1999, Caleffi2024}.
Therefore, generating high-quality quantum states is the cornerstone of advanced quantum information processing~\cite{Bennett1996, Pan2001}.

The quality of a quantum state is degraded by error sources such as environmental noise, imperfections in measurement devices, and statistical fluctuations.
One method to evaluate the quality of a quantum state is quantum state tomography (QST)~\cite{James2001,Takesue2009}.
Using QST, the experimental density matrix can be reconstructed, allowing the state fidelity and trace distance to be calculated~\cite{Nielsen2010}.
However, since the fidelity (or the trace distance) is a scalar, identifying the error sources from this single value alone is unreasonable.
Instead, we use the reconstructed density matrix because it has multiple matrix elements that are better suited for estimating the error sources.
However, this approach is hindered by the contributions of various error sources being superimposed on a single density matrix, which prevents experts from quantifying individual error contributions from the density matrix.
Several methods have been proposed to analyze the quantum operations performed on a target system, including quantum process tomography~\cite{Chuang1997, Lo2020} and gate set tomography~\cite{Blume-Kohout2013, Blume-Kohout2017}, but these methods require more measurements and yield more complex matrices than QST, making it not straightforward to identify error sources.
In contrast, classical shadow~\cite{Huang2020}, self-testing protocols~\cite{Hu2023}, and direct fidelity estimation methods~\cite{Flammia2011} can determine the quality of a quantum state with a limited number of measurements.
Unfortunately, these approaches provide only a quantitative assessment of the overall fidelity or entanglement of the state but do not directly reveal which specific components are responsible for the errors.

\begin{figure}[ht]
 \centering
 \includegraphics[width=0.9\linewidth]{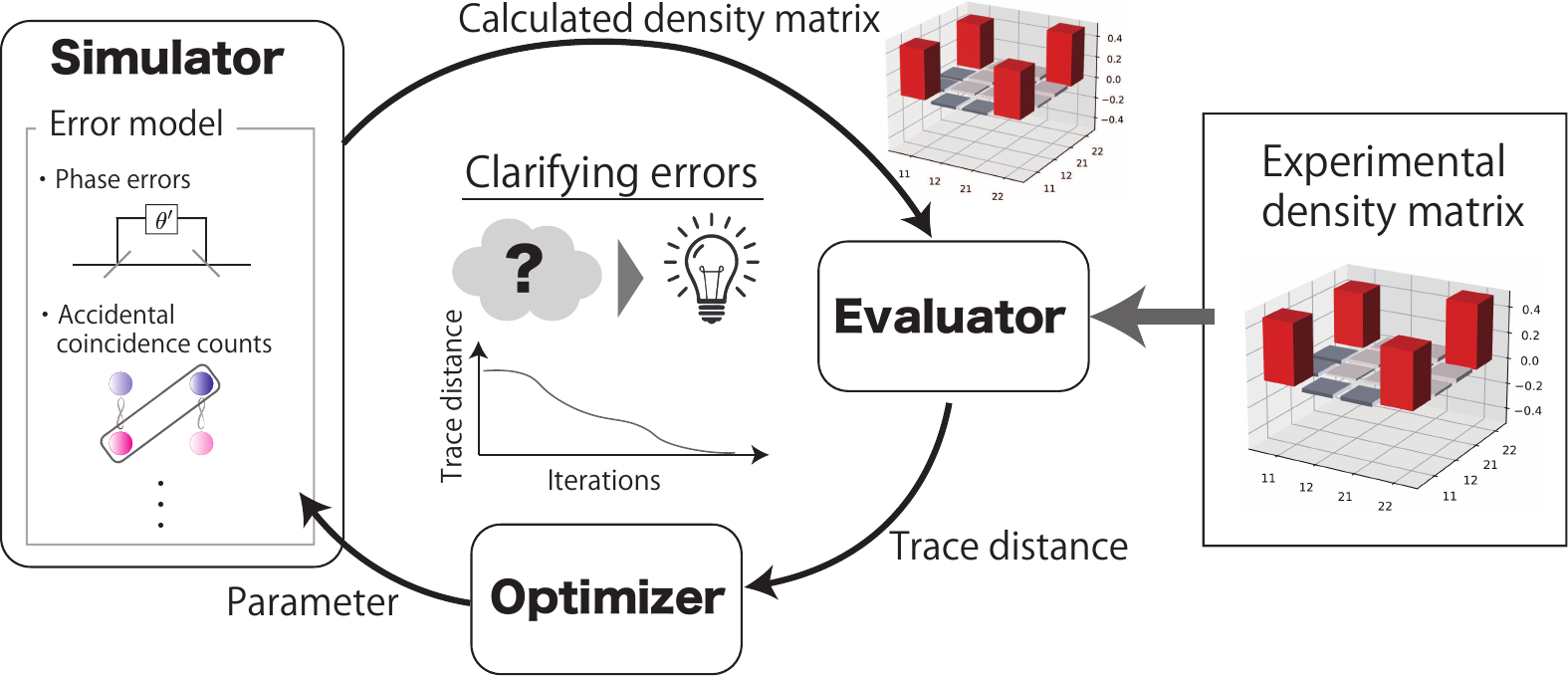}
  \caption{
  Conceptual diagram of MBQEQ workflow.}
\label{fig:concept}
\end{figure}

To overcome these problems, we present an automated quantification method for the error sources that reproduce the experimental density matrix reconstructed by QST.
Our framework is referred to as model-based quantum error quantification (MBQEQ) and is shown schematically in Fig.~\ref{fig:concept}.
It begins by modeling the error sources relevant to a given experiment and simulating the density matrix based on the model parameters.
Next, the evaluation step computes the trace distance to the experimental density matrix, following which the parameter-optimization algorithm suggests values for the model parameters to decrease the trace distance.
The evaluation and parameter-optimization steps are then repeated to determine the values for the model parameters that minimize the trace distance.
This scheme is inspired by a wider body of optimization-based methodologies, such as materials informatics~\cite{Kajita2020, Kikkawa2020, Oba2022, Lee2023}, topology optimization~\cite{Bendsoe2013, Sato2017, Sato2023}, and Hamiltonian learning~\cite{Wang2017, Guimaraes2023}, which are applied in engineering.
Hamiltonian learning estimates the underlying Hamiltonian operator that dictates the system's inherent dynamics and time evolution.
Although it can also be used to quantify errors by specifying how the characterized operations or Hamiltonian parameters deviate from ideal targets, information about the quantified errors is abstract and not necessarily in a form that is understandable by humans.
In contrast, MBQEQ estimates error sources in a human-understandable form, quantifies their contributions to quantum-state deterioration, and guides improvements in high-quality quantum-state generation.
We demonstrate MBQEQ using a fundamental and practical quantum optics experiment involving the generation of time-bin entangled photon pairs.
Because the error model is modular, MBQEQ can easily be transplanted to superconducting qubits, neutral atoms, or solid-state spins by swapping in platform-specific terms. We therefore envision this work as a general-purpose, low-overhead error microscope that complements Hamiltonian learning and other real-time calibration techniques on the road toward fault-tolerant quantum technologies.

The remainder of this paper is organized as follows:
Section~II presents the experimental setup for QST of time-bin entangled photon pairs.
Section~III introduces the evaluator, Sec.~IV describes the error model of the simulator, and Sec.~V explains the optimizer.
In Sec.~VI, the results of the error quantification are presented and discussed, with the experimental results aiming to reduce errors according to the quantification results.
Finally, our conclusions are presented in Sec.~VII.

\section{Experimental setup}
\label{sec:Experimental_setup}

\begin{figure}[ht]
 \centering
 \includegraphics[width=0.99\linewidth]{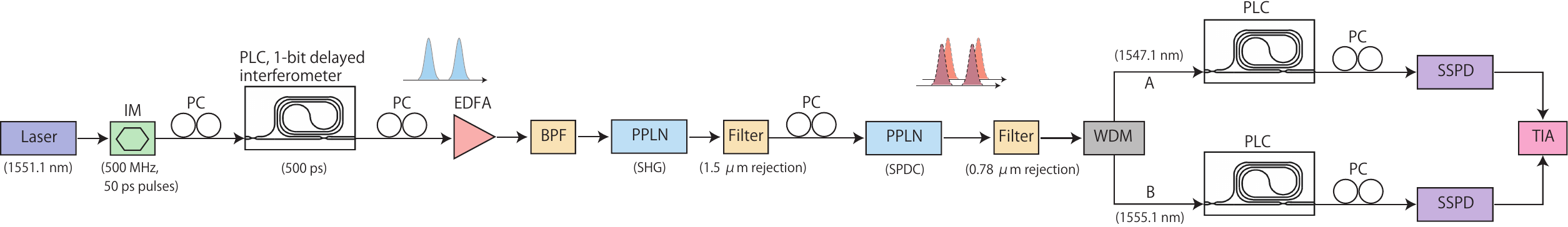}
  \caption{
  Experimental setup for time-bin photon-pair generation.
  IM: intensity modulator. PC: polarization controller. EDFA: erbium-doped fiber amplifier.
  BPF: bandpass filter. PPLN: periodically poled lithium niobate. WDM: wavelength division multiplexing.
  SSPD: superconducting single-photon detector. TIA: time-interval analyzer.}
\label{fig:exp_setup}
\end{figure}

This section describes the details of an experiment used as a concrete example to demonstrate the proposed method. We focus on quantum optics experiments involving the generation of time-bin entangled photon pairs.
Photons can encode quantum information by polarization, frequency, optical path, or time-bin~\cite{Flamini2018, Wang2024}. Although polarization encoding~\cite{Kwiat1995, Takesue2004} is simple to measure, it is susceptible to polarization drift in fibers.
In contrast, time-bin encoding~\cite{Brendel1999, Lo2018, Lo2023, Honjo2008, Inagaki2013, Takesue2015}, where information is encoded in terms of early or late arrival times, is intrinsically robust in long-distance fiber links.
Figure~\ref{fig:exp_setup} shows the experimental setup.
Continuous-wave light at 1551.1 nm emitted from an external-cavity laser diode is converted into a pulse train by an intensity modulator (IM) followed by transformation into double pulses by a 1-bit delayed interferometer implemented by a planar light-wave circuit (PLC)~\cite{Takesue2005}.
The double pulses are amplified by an erbium-doped fiber amplifier (EDFA) and then passed through bandpass filters to suppress the amplified spontaneous-emission noise from the EDFA.
The repetition frequency, pulse interval, and pulse width are 500 MHz, 500 ps, and 50 ps, respectively.
Through second harmonic generation (SHG) inside the first periodically poled lithium niobate (PPLN) waveguide, the 775.55-nm light is generated.
The double pulses serve as pump pulses for spontaneous parametric down-conversion (SPDC) inside the second PPLN waveguide, where a time-bin entangled photon pair is generated with wavelengths of 1547.1 and 1555.1 nm for the two photons.
The photon pair is split by wavelength division multiplexing (WDM) into separate optical paths with a 1-nm bandwidth, each passing through its own 1-bit delayed interferometer before being detected by a superconducting single-photon detector.
The detection efficiencies of the detectors for photons at wavelengths of 1547.1 and 1555.1 nm are 60\% and 27\%, respectively.
The dark count rate is less than 10 cps for both detectors.
The photon detection timing is analyzed by a time-interval analyzer (TIA) for the coincidence count measurement.
The time window is 200 ps for each pulse.
Based on the two-photon coincidence count data, QST is performed to reconstruct the density matrix (see Appendix~\ref{sec:Quantum_state_tomography}).

We denote the state where a photon exists in the early (late) time bin as $\ket{1}$ ($\ket{2}$). The two-photon state of the ideal time-bin-entangled photon pair is given by
\begin{align}
\ket{\Phi}_{\rm AB} = \frac{1}{\sqrt{2}}(\ket{11}_{\rm AB} + \ket{22}_{\rm AB}),
\label{eq:ideal_state}
\end{align}
and its density matrix is $\rho_{\rm ideal} = \ket{\Phi}_{\rm AB}\!\bra{\Phi}$.
Subscripts A and B label the measurement channels split in the WDM as shown in Fig.~\ref{fig:exp_setup}, where the wavelengths are 1547.1 and 1555.1 nm, respectively.
On the measurement side, via the 1-bit delayed interferometers, projective measurements are performed on the states $\ket{1}_{\rm A(B)}$, $\ket{2}_{\rm A(B)}$, and their superpositions
\begin{align}
\ket{+}_{\rm A(B)} &= \frac{1}{\sqrt{2}} \left( \ket{1}_{\rm A(B)} + e^{-i \theta^+_{\rm A(B)}}\ket{2}_{\rm A(B)} \right), \\
\ket{L}_{\rm A(B)} &= \frac{1}{\sqrt{2}} \left( \ket{1}_{\rm A(B)} + e^{-i \theta^L_{\rm A(B)}}\ket{2}_{\rm A(B)} \right),
\end{align}
where the phases $\theta^+_{\rm A(B)} = 0$ and $\theta^L_{\rm A(B)} = - \frac{\pi}{2}$ are added by adjusting the PLC temperature.
Before performing QST, the temperatures corresponding to $\theta^+_{\rm A}$ and $\theta^+_{\rm B}$ are calibrated to obtain the maximum coincidence.
The temperatures corresponding to $\theta^L_{\rm A} = - {\pi}/{2}$ and $\theta^L_{\rm B} = - {\pi}/{2}$ are calibrated by fitting a coincidence curve around this maximum point.
Temperatures are set to $66.16\,^{\circ}$C ($36.41\,^{\circ}$C) and $65.92\,^{\circ}$C ($36.15\,^{\circ}$C) for the $\ket{+}_{\rm A(B)}$ and $\ket{L}_{\rm A(B)}$ states, respectively.
By linear QST~\cite{Takesue2009}, the experimental density matrix ($\rho_{\rm exp}$), which has 16 matrix elements, is reconstructed from the coincidence counts of the projective measurements onto 16 states, $\{\ket{1}_A, \ket{2}_A, \ket{+}_A, \ket{L}_A\} \otimes \{\ket{1}_B, \ket{2}_B, \ket{+}_B, \ket{L}_B\}$ (see Appendix~\ref{sec:Quantum_state_tomography} for the details).

\begin{figure}[ht]
 \centering
 \includegraphics[width=0.5\linewidth]{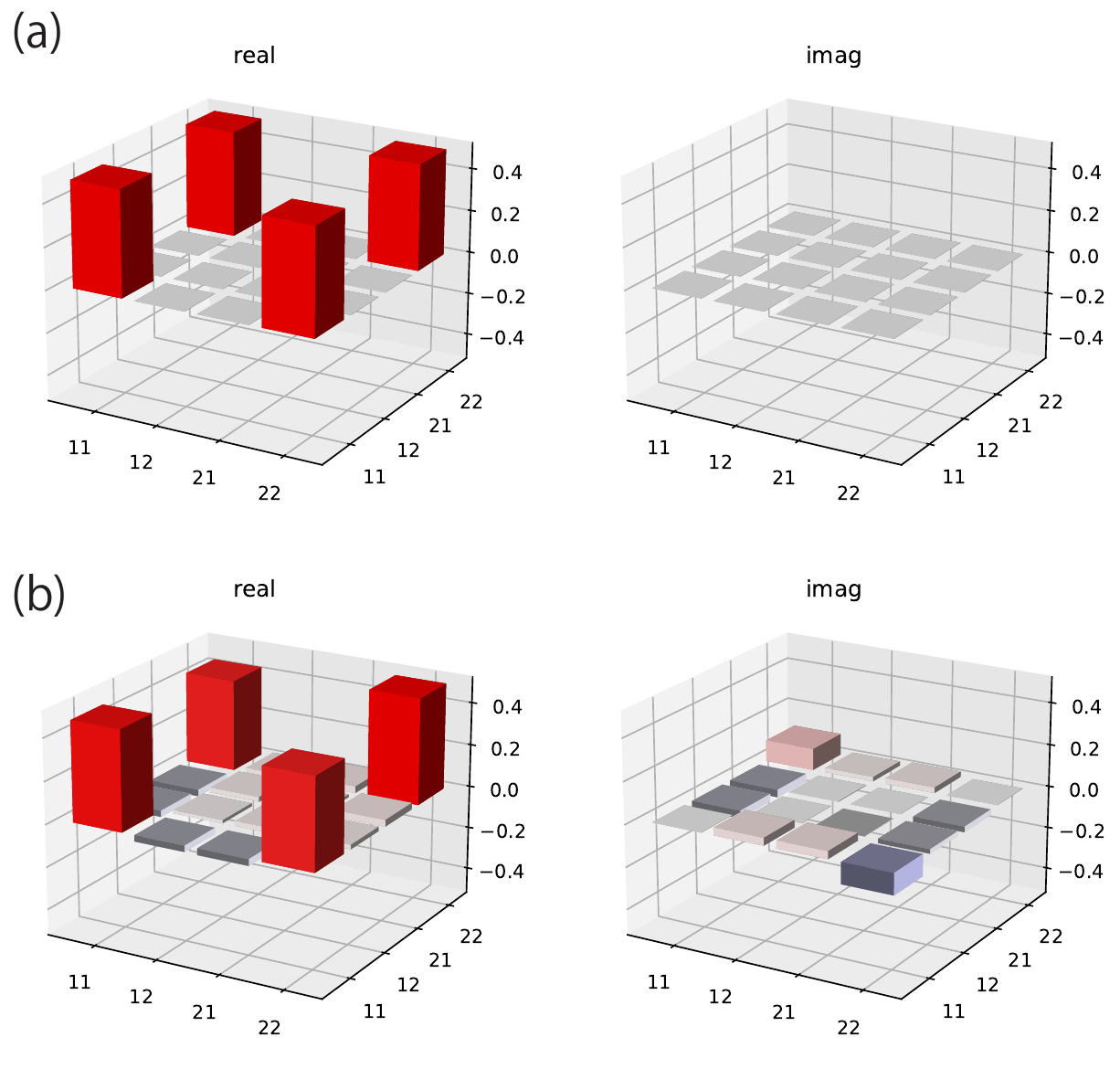}
  \caption{
  Density matrices of (a) ideal state and (b) experimental result.
  The left (right) panel shows the real (imaginary) part.}
\label{fig:rho}
\end{figure}

Figures~\ref{fig:rho}(a) and (b) show the density matrices of the ideal state ($\rho_{\rm ideal}$) and the experimental result, respectively. 
In the ideal state, peaks appear at the four corners of the real part, whereas the experimental result exhibits additional components that degrade the fidelity.
Their fidelity,
\begin{align}
F(\rho_{\rm ideal}, \rho_{\rm exp}) = \bra{\Phi}_{\rm AB}\rho_{\rm exp}\ket{\Phi}_{\rm AB},
\label{eq:Fidelity}
\end{align}
is 91.6\%.

\section{Evaluator}
\label{sec:Evaluator}

The experimental density matrix reconstructed by linear QST is often non-positive definite, meaning that it has at least one negative eigenvalue (see Appendix~\ref{sec:Eigenvalue_decomposition} for details of the eigenvalue decomposition), which can be caused by errors such as experimental imperfections or statistical fluctuations.
Although the non-positive definite component is usually considered a nuisance, it is informative for quantifying the error sources in MBQEQ.
To quantify the error, including the non-positive definite components, we use a suitable distance measure, namely, the trace distance.
Therefore, the evaluator calculates the trace distance between the aforementioned experimental density matrix and the simulated density matrix described in the next section.
The trace distance of two arbitrary density matrices ($\rho, \sigma$) is defined as follows:
\begin{align}
D(\rho, \sigma) = \frac{1}{2}{\rm Tr}|\rho - \sigma|.
\end{align}

\section{Simulator}
\label{sec:Simulator}

The simulator simulates the reconstructed density matrix via linear QST based on the parametrized error model.
The details are given in the following subsections.

\subsection{Accidental coincidences}
\label{sec:Accidental_coincidences}

Our experiment accidentally generates more than one photon pair according to a Poisson distribution because the optical bandwidths of the SPDC photons are much larger than the inverse of the temporal pulse width and multiple pairs are sufficiently distinguishable~\cite{Takesue2010}.
These multiple photon pairs lead to accidental coincidence counts (i.e., coincidences with uncorrelated photons).
Such accidental coincidences introduce uniform random errors in the QST measurements, degrading the fidelity of the obtained state.
We introduce this error as the depolarizing channel $\mathcal{E}$~\cite{Nielsen2010}.
For any density matrix $\rho$ of a two-qubit system, $\mathcal{E}$ performs the operation
\begin{align}
\mathcal{E}(\rho) &= (1-\eta)\rho + \eta\frac{I_2}{4},
\end{align}
where $I_2$ is the identity operator on the two-qubit Hilbert space.
When $\eta = 1$, the two-qubit system becomes the maximally mixed state $I_2/4$.
Appendix~\ref{sec:Details_of_accidental_coincidence_error} justifies modeling the experiment in this way.

\subsection{Phase error in measurement basis}
\label{sec:Phase_error_in_measurement_basis}

When projecting onto the states $\ket{+}_{\rm A(B)}$ and $\ket{L}_{\rm A(B)}$, a phase shift on the long optical path in the 1-bit delayed interferometer is introduced by adjusting the PLC temperature.
However, deviations from the intended phase shift can occur accidentally due to the calibration error and errors in the temperature controller.
We introduce phase errors in the measurement bases through four parameters
$\theta^{'+}_{\rm A(B)}, \theta^{'L}_{\rm A(B)}$ as follows:
\begin{align}
\theta^+_{\rm A(B)} &= 0 + \theta^{'+}_{\rm A(B)}, \\
\theta^L_{\rm A(B)} &= - \frac{\pi}{2} + \theta^{'L}_{\rm A(B)}.
\end{align}
These deviations of the states affect the density matrix reconstructed by QST (see Appendix~\ref{sec:Quantum_state_tomography}).
The summation of the phases of two measurements only affects the coincidence counts of the entangled state, so combinations of the phase errors $\{\theta^{'+}_{\rm A}+\theta^{'+}_{\rm B},\ \theta^{'+}_{\rm A}+\theta^{'L}_{\rm B},\ \theta^{'L}_{\rm A}+\theta^{'+}_{\rm B},\ \theta^{'L}_{\rm A}+\theta^{'L}_{\rm B} \}$ constitute the net errors that affect the reconstructed density matrix.

\subsection{Intensity error in measurement basis}
\label{sec:Intensity_error_in_measurement_basis}

In addition to the phase error in the measurement basis, an intensity error can be induced by an imperfect PLC (e.g., a difference in loss between the short and long arms or an imperfect half beamsplitter).
We introduce intensity errors in the measurement bases through two parameters $p_{\rm A}$ and $p_{\rm B}$ as follows:
\begin{align}
\ket{+}_{\rm A(B)} &= \sqrt{p_{\rm A(B)}} \ket{1}_{\rm A(B)} + e^{-i \theta^+_{\rm A(B)}} \sqrt{1-p_{\rm A(B)}} \ket{2}_{\rm A(B)}, \\
\ket{L}_{\rm A(B)} &= \sqrt{p_{\rm A(B)}} \ket{1}_{\rm A(B)} + e^{-i \theta^L_{\rm A(B)}} \sqrt{1-p_{\rm A(B)}} \ket{2}_{\rm A(B)}.
\end{align}

\subsection{Intensity asymmetry}
\label{sec:Intensity_asymmetry}

Although the ideal state produces equal intensities in the early and late time bins, the intensities can vary in actual experiments for the same reason as the intensity error in the measurement basis, namely, because the PLC prepares the double pulses.
We introduce time-bin intensity asymmetry via parameter $p$ as follows:
\begin{align}
\ket{\Phi'}_{\rm AB} =
\sqrt{p} \ket{11}_{\rm AB}
+ \sqrt{1-p} \ket{22}_{\rm AB}.
\end{align}
In the ideal state, $p = 0.5$.
Departing from this value varies the intensity ratio between the $\ket{11}$ and $\ket{22}$ components.

\subsection{Statistical fluctuations}
\label{sec:Statistical_fluctuations}

Given that the photon counts follow the Poisson distribution, it has a statistical fluctuation $\sqrt{n_\nu}$ if the obtained coincidence count is $n_\nu$ for the $\nu$th projected state ($\nu \in \{1,2,\dots,16\}$.
Thus, the coincidence probability, defined as $s_\nu = n_\nu /C$ with a normalization constant $C$, has width $\sigma_\nu = \sqrt{n_\nu}/C$.
We introduce 16 parameters denoted $\delta_\nu \in [-\sigma_\nu, \sigma_\nu]$ for each $s_\nu$ to represent the statistical fluctuations for the results of the simulated measurement (see Appendix~\ref{sec:Quantum_state_tomography}).

\subsection{Relative phase}
\label{sec:Relative_phase}

An error in the PLC temperature control introduces a relative phase error between the double pulses, which is directly transferred to the state of the photon pair generated by SPDC, thereby reducing the fidelity with respect to the ideal state.
We introduce the relative phase error through the parameter $\theta_{22}$ as follows:
\begin{align}
\ket{\Phi'}_{\rm AB} =
\frac{1}{\sqrt{2}} \left( \ket{11}_{\rm AB}
+ e^{i \theta_{22}}\ket{22}_{\rm AB} \right).
\end{align}

\subsection{Correlation of photon pair}
\label{sec:Correlation_of_photon_pair}

In general, energy conservation induces frequency correlations between photon pairs generated by SPDC.
When Fourier transformed into real space, these frequency correlations affect the width of the photon wave packets.
The broadened pulses may have tails that extend into adjacent time bins, potentially introducing errors in the time-bin entangled state.
In our simulation, the two-photon wave function in real space is modeled by a Gaussian function.
We introduce a model parameter representing the photon-pair correlation as $r_{\rm corr}$, which is defined as the ratio of the long-to-short axes of a two-dimensional Gaussian tilted by $45^\circ$ (see Appendix~\ref{sec:Probability_distribution_of_two-photon_state}).

\subsection{Summary of error model}
\label{sec:Summary_of_error_model}

By adding the modeled errors to the ideal state defined in Eq.~\eqref{eq:ideal_state}, the state can be written as
\begin{align}
\ket{\Phi'}_{\rm AB} =
\sqrt{p} \ket{11}_{\rm AB}
+ e^{i \theta_{22}} \sqrt{1-p} \ket{22}_{\rm AB}.
\label{eq:Phi'_all}
\end{align}
Note that $p = 0.5$ and $\theta_{22} =  0$ reproduce the ideal state.
In addition to Eq.~\eqref{eq:Phi'_all}, we introduce the frequency correlation of two photons parametrized by $r_{\rm corr}$, which affects the pulse width of the two-photon state.
However, the implementation of $r_{\rm corr}$ is not straightforward because the pulse-shape information is more complex and requires more degrees of freedom than Eq.~\eqref{eq:Phi'_all}.
To include such fine-grained information with the parameters in Eq.~\eqref{eq:Phi'_all}, we use a quantum optics simulator~\cite{Oba2024} (see Appendix~\ref{sec:Probability_distribution_of_two-photon_state} for the details).

The density matrix $\rho_{\rm sim}$ calculated from the error model is then given by
\begin{align}
\rho_{\rm sim} = \mathcal{T}' \left[\mathcal{E}\left(\ket{\Phi'}_{\rm AB}\!\bra{\Phi'}\right)\right],
\end{align}
where $\mathcal{T}'$ denotes the QST operation, which accounts for phase and intensity errors in the measurement bases as well as statistical fluctuations and is parametrized by $\{\theta^{'+}_{\rm A},\ \theta^{'L}_{\rm A},\ \theta^{'+}_{\rm B},\ \theta^{'L}_{\rm B},\ p_{\rm A},\ p_{\rm B},\ \delta_\nu\}$.
Note that the operation $\mathcal{T}'$ is not necessarily a completely positive and trace-preserving map.
$\mathcal{E}$ denotes a depolarizing channel parametrized by $\eta$.

\section{Optimizer}
\label{sec:Optimizer}

The optimizer iteratively optimizes the model parameters to minimize a given cost function.
For parameter optimization, we use Powell's method~\cite{Powell1964} as implemented in the Python library \texttt{SciPy} (https://scipy.org).
For the cost function, we use the trace distance~\cite{Nielsen2010} between the experimental density matrix and the calculated density matrix,
\begin{align}
D(\rho_{\rm exp}, \rho_{\rm sim}) = \frac{1}{2}{\rm Tr}|\rho_{\rm exp} - \rho_{\rm sim}|.
\end{align}
The boundaries of the model parameters are set as follows:
$r_{\rm corr} \in [1, 3]$,
$\theta_{22} \in [-\pi/2, \pi/2]$,
$p \in [0.2, 0.8]$,
$p_{\rm A(B)} \in [0.2, 0.8]$
$\theta^{'+}_{\rm A(B)} \in [-\pi/2, \pi/2]$,
$\theta^{'L}_{\rm A(B)} \in [-\pi/2, \pi/2]$,
$\eta \in [0, 1]$, and
$\delta_\nu \in [-\sigma_\nu, \sigma_\nu]$.
The initial values of the parameters are set to those of the ideal state: $r_{\rm corr}=1$, $p=p_{\rm A(B)}=0.5$, and $0$ for the rest.
To avoid excessive fitting by $\delta_\nu$, we first estimate all the model parameters except $\delta_\nu$.
Then, $\delta_\nu$ is estimated while fixing the other model parameters to those obtained in the first step.
This two-step optimization process took approximately 30 minutes on a single core of an ordinary personal computer (M2 MacBook Air with 16 GB memory).
Separately, in order to assess the stability of the method, we ran 100 optimizations with different initial guesses sampled from a uniform distribution and confirmed that all runs converged to similar parameter values (see Appendix~\ref{sec:Stability_analysis} for details).

\section{Results and discussion}
\label{sec:Results_and_discussion}

\subsection{Quantification of intentional accidental coincidence errors}
\label{sec:Quantification_of_intentional_accidental_coincidence_errors}

\begin{figure}[ht]
 \centering
 \includegraphics[width=0.99\linewidth]{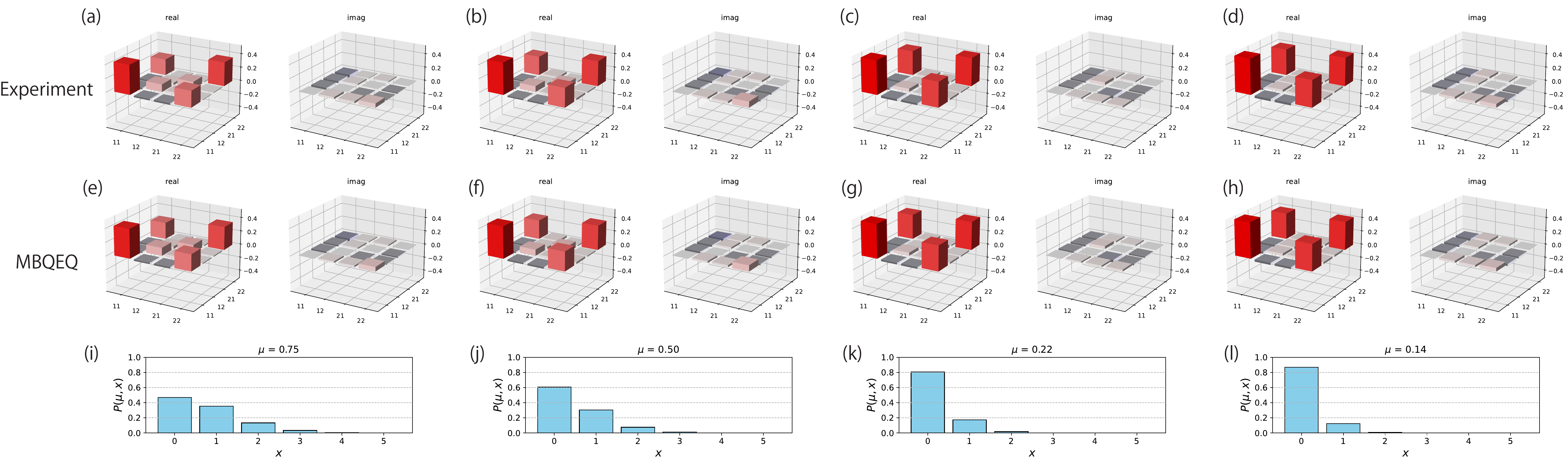}
  \caption{
  (a)-(d) Density matrices obtained from experiments with accidental coincidence errors.
  The pump power is increasingly attenuated from the leftmost panel to the rightmost panel.
  (e)-(h) Density matrices obtained from MBQEQ.
  (i)-(l) Poisson distribution of the number of photon pairs. The mean photon-pair number is calculated from the expected visibility (see Appendix~\ref{sec:Details_of_accidental_coincidence_error} for details).}
\label{fig:rho_eta_error}
\end{figure}

To determine whether the proposed method can identify an imposed error, we intentionally introduce an accidental coincidence error and subsequently perform MBQEQ.
Accidental coincidences occur when the SPDC generates multiple photon pairs.
Since the number of photon pairs generated increases with pump power, a higher pump intensity leads to a larger mean photon-pair number, so the accidental coincidence error should increase.
By adjusting the pump power with an attenuator, we perform QST experiments and MBQEQ at several intensities.
The results appear in Fig.~\ref{fig:rho_eta_error}, and Table~\ref{tab:opt_val_eta_err} lists the parameter values obtained from MBQEQ.
The value of the model parameter $\eta$ obtained from MBQEQ is consistent with that estimated experimentally ($\eta_{\rm exp}$), whereas the other parameters are not significantly affected (see Appendix~\ref{sec:Details_of_accidental_coincidence_error} for the details of $\eta_{\rm exp}$).
The results demonstrate that MBQEQ identifies and quantifies accidental coincidence errors intentionally introduced in the experiment.

\begin{table}[ht]
 \caption{Values of error model parameters determined by MBQEQ with intentional accidental coincidence errors. The degrees are given in parentheses.
 $\eta_{\rm exp}$ shows the value of the model parameter estimated from each experiment, and $\mu$ shows the corresponding mean photon-pair number. The statistical fluctuation parameters are omitted for brevity.}
 \label{tab:opt_val_eta_err}
 \centering
  \begin{tabular}{|c|c|c|c|c|c|c|c|c|c||c|c|}
   \hline
   $r_{\rm corr}$& $\theta_{22}$ & $p$ & $p_{\rm A}$ & $p_{\rm B}$ &
   $\theta^{'+}_{\rm A}+\theta^{'+}_{\rm B}$ & $\theta^{'+}_{\rm A}+\theta^{'L}_{\rm B}$ &
   $\theta^{'L}_{\rm A}+\theta^{'+}_{\rm B}$ & $\theta^{'L}_{\rm A}+\theta^{'L}_{\rm B}$ & $\eta$ &
   $\eta_{\rm exp}$ & $\mu$ \\
   \hline
   2.2 & 0.001 & 0.52 & 0.47 & 0.47 & 0.05 & $-$0.33 & $-$0.20 & $-$0.58 & {\bf 0.45} & 0.43 & 0.75\\
   & ($0.06^\circ$) & & & & ($ 2.7^\circ$) & ($ -18.9^\circ$) & ($ -11.6^\circ$) & ($ -33.2^\circ$) & & & \\
   \hline  
   2.3 & 0.002 & 0.51 & 0.46 & 0.47 & $-$0.02 & $-$0.38 & $-$0.24 & $-$0.59 & {\bf 0.36} & 0.34 & 0.50\\
   & ($0.1^\circ$) & & & & ($ -1.2^\circ$) & ($ -21.5^\circ$) & ($ -13.5^\circ$) & ($ -33.8^\circ$) & & & \\
   \hline  
   2.0 & 0.01 & 0.50 & 0.47 & 0.48 & 0.37 & $-$0.17 & 0.01 & $-$0.53 & {\bf 0.18} & 0.18 & 0.22\\
   & ($0.7^\circ$) & & & & ($ 21.3^\circ$) & ($ -10.0^\circ$) & ($ 0.8^\circ$) & ($ -30.4^\circ$) & & & \\
   \hline  
   2.2 & $-$0.002 & 0.50 & 0.47 & 0.47 & 0.34 & $-$0.18 & $-$0.07 & $-$0.59 & {\bf 0.12} & 0.12 & 0.14\\
   & ($-0.1^\circ$) & & & & ($ 19.4^\circ$) & ($ -10.3^\circ$) & ($ -4.0^\circ$) & ($ -33.7^\circ$) & & & \\
   \hline
  \end{tabular}
\end{table}

\subsection{Quantification of intentional phase errors in measurement bases}
\label{sec:Quantification_of_intentional_phase_errors_in_measurement_bases}

\begin{figure}[ht]
 \centering
 \includegraphics[width=0.99\linewidth]{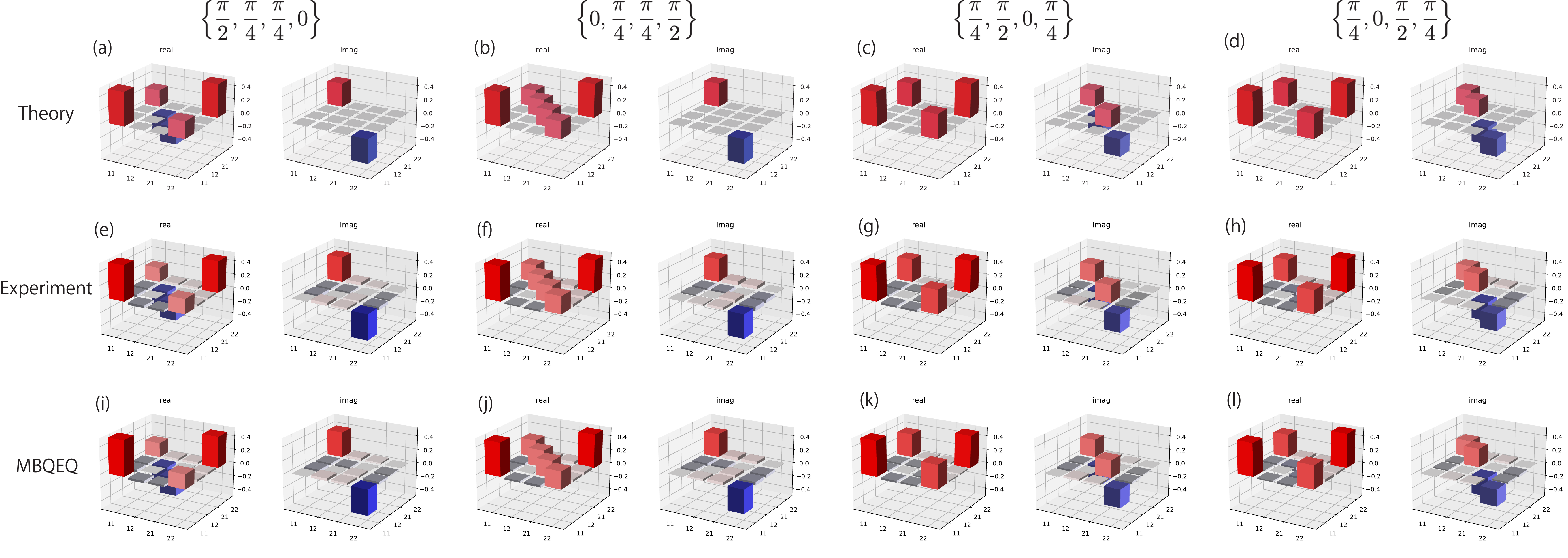}
  \caption{
  (a)-(d) Density matrices obtained from theoretical calculations with intentional phase errors.
  The intended phase errors $\{\theta^{'+}_{\rm A}+\theta^{'+}_{\rm B},\ \theta^{'+}_{\rm A}+\theta^{'L}_{\rm B},\ \theta^{'L}_{\rm A}+\theta^{'+}_{\rm B},\ \theta^{'L}_{\rm A}+\theta^{'L}_{\rm B} \}$ are shown above the panels.
  (e)-(h) Density matrices obtained experimentally.
  (i)-(l) Density matrices obtained from MBQEQ.}
\label{fig:rho_theta_error}
\end{figure}

In addition, by deliberately shifting the PLC temperature during the QST, we intentionally introduce phase errors in the measurement bases and subsequently perform MBQEQ.
The temperatures are denoted $\{T_0,\ T_{-\pi/2}\}$ for the phases $\{0,\ -\pi/2\}$, respectively.
Because the induced phase $\theta\propto T$, shifting the temperature by $\Delta T_{\pi/4} = |T_0 - T_{-\pi/2}|/2$ suffices to shift the phase by $\pi/4$.
Therefore, the $\pi/4$-shifted temperatures become $\{T_0 +\Delta T_{\pi/4},\ T_{-\pi/2}+\Delta T_{\pi/4}\}$.
Among the four phases $\{\theta^{+}_{\rm A},\ \theta^{L}_{\rm A},\ \theta^{+}_{\rm B},\ \theta^{L}_{\rm B}\}$ in the measurement bases, we shift two of them by $\pi/4$ and perform the QST experiment.
The MBQEQ is also performed for each experimental density matrix.
The results are shown in Fig.~\ref{fig:rho_theta_error} and Table~\ref{tab:opt_val_theta_err}.
Given that the temperature controller used in this experiment has a precision of approximately $0.026\,^\circ$C around $66\,^\circ$C (the temperature used), the phase may deviate by roughly $9^\circ$ from the intended value.
Table~\ref{tab:opt_val_theta_err} shows that the results of MBQEQ are consistent with the intended phase errors, further supporting the reliability of MBQEQ.

\begin{table}[ht]
 \caption{Values of error model parameters determined by MBQEQ with intentional phase errors in the measurement bases. The degrees are given in parentheses.
 The 11th column shows the intended phase errors $\{\theta^{'+}_{\rm A}+\theta^{'+}_{\rm B},\ \theta^{'+}_{\rm A}+\theta^{'L}_{\rm B},\ \theta^{'L}_{\rm A}+\theta^{'+}_{\rm B},\ \theta^{'L}_{\rm A}+\theta^{'L}_{\rm B} \}$. The statistical fluctuation parameters are omitted for brevity.}
 \label{tab:opt_val_theta_err}
 \centering
  \begin{tabular}{|c|c|c|c|c|c|c|c|c|c||c|}
   \hline
   $r_{\rm corr}$ & $\theta_{22}$ & $p$ & $p_{\rm A}$ & $p_{\rm B}$ &
   $\theta^{'+}_{\rm A}+\theta^{'+}_{\rm B}$ & $\theta^{'+}_{\rm A}+\theta^{'L}_{\rm B}$ &
   $\theta^{'L}_{\rm A}+\theta^{'+}_{\rm B}$ & $\theta^{'L}_{\rm A}+\theta^{'L}_{\rm B}$ & $\eta$ &
   Intended phase error \\
   \hline
   1.8 & $-$0.005 & 0.47 & 0.47 & 0.47 & {\bf 1.70} & {\bf 0.92} & {\bf 0.77}  & {\bf 0.002} & 0.01 &
   $\{ \frac{\pi}{2}, \frac{\pi}{4}, \frac{\pi}{4}, 0\}$ \\
   & ($-0.3^\circ$) & & & & ($ {\bf 97.2^\circ}$) & ($ {\bf 52.9^\circ}$) & ($ {\bf 44.4^\circ}$) & ($ {\bf 0.1^\circ}$) & & \\
   \hline
   2.3 & $-$0.002 & 0.42 & 0.46 & 0.46 & {\bf 0.04} & {\bf 0.87} & {\bf 0.70} & {\bf 1.54} & 0.02 &
   $\{ 0, \frac{\pi}{4}, \frac{\pi}{4}, \frac{\pi}{2} \}$ \\
   & ($-0.1^\circ$) & & & & ($ {\bf 2.1^\circ}$) & ($ {\bf 49.7^\circ}$) & ($ {\bf 40.1^\circ}$) & ($ {\bf 87.8^\circ}$) & & \\
   \hline
   1.8 & $-$0.002 & 0.48 & 0.48 & 0.48 & {\bf 0.87} & {\bf 1.60} & {\bf 0.006} & {\bf 0.74} & 0.01 &
   $\{ \frac{\pi}{4}, \frac{\pi}{2}, 0, \frac{\pi}{4} \}$ \\
   & ($-0.1^\circ$) & & & & ($ {\bf 49.8^\circ}$) & ($ {\bf 91.7^\circ}$) & ($ {\bf 0.3^\circ}$) & ($ {\bf 42.3^\circ}$) & & \\
   \hline
   2.1 & 0.002 & 0.44 & 0.47 & 0.47 & {\bf 0.98} & {\bf $-$0.05} & {\bf 1.64} & {\bf 0.61} & 0.01 &
   $\{ \frac{\pi}{4}, 0, \frac{\pi}{2}, \frac{\pi}{4} \}$ \\
   & ($0.1^\circ$) & & & & ($ {\bf 56.2^\circ}$) & ($ {\bf -2.9^\circ}$) & ($ {\bf 94.2^\circ}$) & ($ {\bf 35.1^\circ}$) & & \\
   \hline  \end{tabular}
\end{table}

\subsection{Identification of high-impact error sources by ablation study}
\label{sec:Identification_of_high-impact_error_sources_by_ablation_study}

\begin{figure}[ht]
 \centering
 \includegraphics[width=0.9\linewidth]{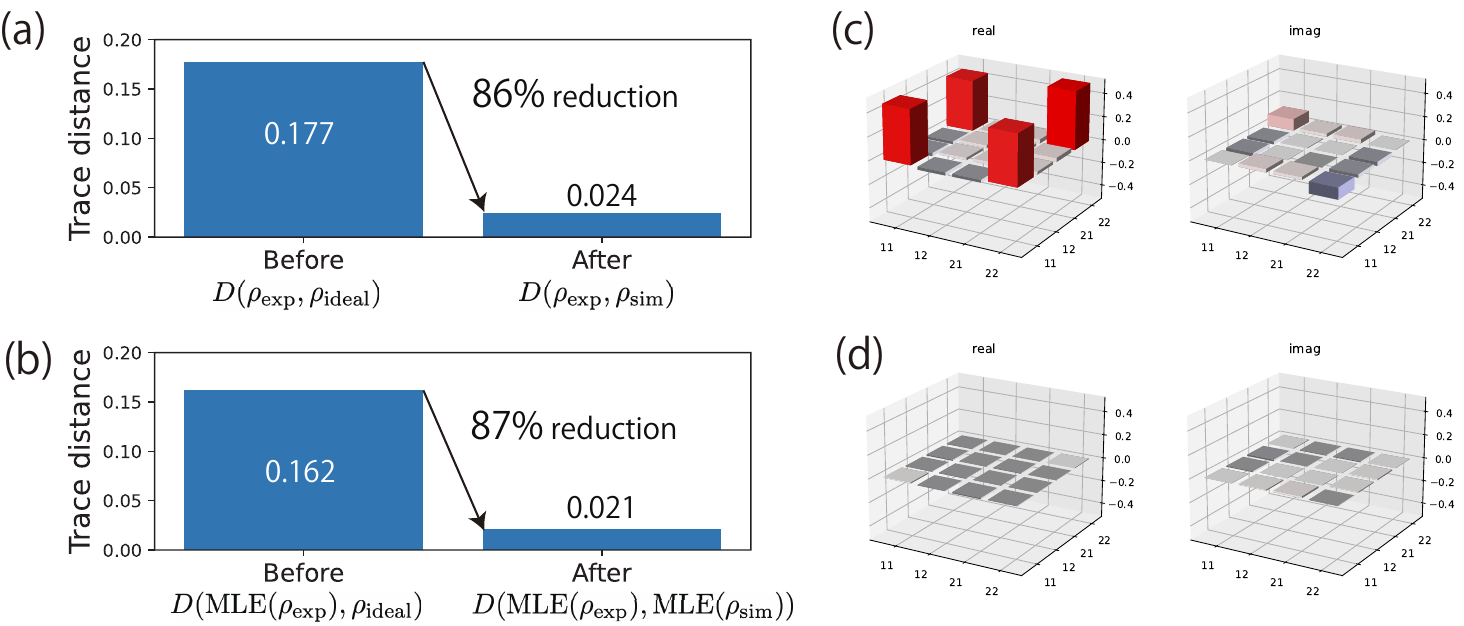}
  \caption{
  (a) Trace distances between experimental and simulated density matrices before and after MBQEQ.
  (b) Same as panel (a) but showing the result after applying maximum likelihood estimation (MLE) to eliminate non-positive definite error components.
  (c) Density matrix after MBQEQ.
  (d) Residual after subtracting the density matrix in panel (c) from the experimental density matrix (Fig.~\ref{fig:rho}(b)).}
\label{fig:d_trace}
\end{figure}

After confirming the performance of the proposed method, we analyze the baseline data ($\rho_{\rm exp}$ shown in Fig.~\ref{fig:rho}(b)) to demonstrate its practical utility.
Figure~\ref{fig:d_trace}(a) shows the trace distances before and after MBQEQ. 
The value labeled ``Before" is the trace distance between the ideal state and the experimental result, and the value labeled ``After" is the trace distance between the simulated state and the experimental result.
By optimizing the error model parameters, the error sources are added to the ideal state, and the trace distance to the experimental result decreases significantly from 0.177 to 0.024.
This result indicates that 86\% of the errors are estimated automatically by the proposed method.
Figures~\ref{fig:d_trace}(c) and (d) show the density matrix after parameter optimization and its difference from the experimental result (see also Fig.~\ref{fig:rho}(b)).
The main peaks are accurately reproduced, although small residuals persist.
We also compare the eigenvalues and eigenstates, confirming that they are consistent (including the negative eigenvalue component; see Appendix~\ref{sec:Eigenvalue_decomposition} for the details).
The remaining errors may stem from detector imperfections and external noise~\cite{Hradil1997}, which are not implemented in this study.
Note, however, that they account for less than 1\% of the fidelity degradation, as shown in the next paragraph.
The estimated parameters are listed in Table~\ref{tab:opt_val}.

\begin{table}[ht]
 \caption{Model parameters estimated by MBQEQ. The degrees are given in parentheses.
 The corresponding projected states are written below the statistical fluctuation parameters.}
 \label{tab:opt_val}
 \centering
 \resizebox{\textwidth}{!}{%
  \begin{tabular}{|c|c|c|c|c|c|c|c|c|c|c|c|c|c|c|c|}
   \cline{1-10}
   $r_{\rm corr}$& $\theta_{22}$ & $p$ & $p_{\rm A}$ & $p_{\rm B}$ &
   $\theta^{'+}_{\rm A} + \theta^{'+}_{\rm B}$ & $\theta^{'+}_{\rm A} + \theta^{'L}_{\rm B}$ &
   $\theta^{'L}_{\rm A} + \theta^{'+}_{\rm B}$ & $\theta^{'L}_{\rm A} + \theta^{'L}_{\rm B}$ & $\eta$ \\
   \cline{1-10}
   1.3 & 0.001 & 0.43 & 0.48 & 0.47 & $-$0.04 & 0.20 & 0.23 & 0.46 & 0.075 \\
   & ($0.06^\circ$) & & & & ($ -2.1^\circ$) & ($ 11.2^\circ$) & ($ 12.9^\circ$) & ($ 26.2^\circ$) & \\
   \hline \hline
   $\delta_1$ & $\delta_2$ & $\delta_3$ & $\delta_4$ & $\delta_5$ &
   $\delta_6$ & $\delta_7$ & $\delta_8$ & $\delta_9$ & $\delta_{10}$ &
   $\delta_{11}$ & $\delta_{12}$ & $\delta_{13}$ & $\delta_{14}$ & $\delta_{15}$ & $\delta_{16}$ \\
   $\ket{11}$ & $\ket{12}$ & $\ket{1+}$ & $\ket{1L}$ & $\ket{21}$ &
   $\ket{22}$ & $\ket{2+}$ & $\ket{2L}$ & $\ket{+1}$ & $\ket{+2}$ &
   $\ket{++}$ & $\ket{+L}$ & $\ket{L1}$ & $\ket{L2}$ & $\ket{L+}$ & $\ket{LL}$ \\
   \hline
   $-4\!\times\!10^{-6}\,$\% & $2\!\times\!10^{-6}\,$\% & 0.17\% & $-$0.21\% &
   $1\!\times\!10^{-5}\,$\% & $-2\!\times\!10^{-5}\,$\% & 0.04\% & 0.12\%  &
   $-$0.13\% & $-$0.17\% & $-$0.43\% & $-$0.24\% &
   $-$0.29\% & 0.19\% & $-$0.33\% & $-$0.11\% \\
   \hline       
  \end{tabular}
  }
\end{table}

To improve state fidelity, we use the information of the quantified error sources to identify the errors that significantly improved fidelity.
However, the non-positive definite matrix makes the fidelity exceed unity, which hinders the precise assessment of the improvement in fidelity.
Therefore, in the following discussion, we remove the non-positive definite components from the density matrix by performing maximum likelihood estimation (MLE)~\cite{James2001, Takesue2009}, which is a common approach for obtaining a physical (positive definite) density matrix (see Appendix~\ref{sec:Maximum_likelihood_estimation} for the details).
Here, we denote the MLE-applied state, which is a positive definite density matrix, by MLE($\rho$), where $\rho$ is a quantum density matrix.
Note that, although we apply MLE to ensure a maximum fidelity of unity, the resulting discussion is essentially the same as that without MLE (see Fig.~\ref{fig:d_trace}(b) and note that the trace distance after MBQEQ remains unchanged from that in Fig.~\ref{fig:d_trace}(a)).

\begin{figure}[ht]
 \centering
 \includegraphics[width=0.7\linewidth]{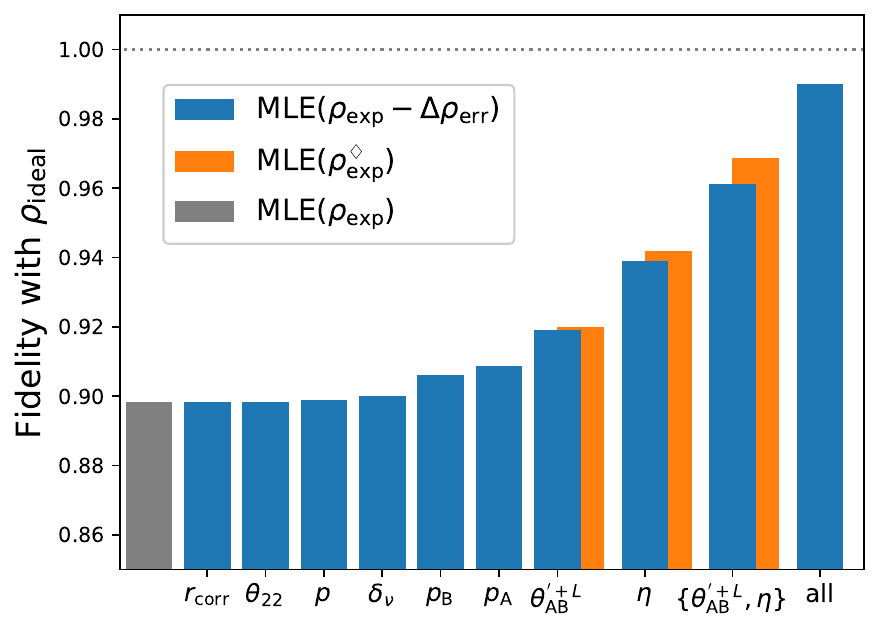}
  \caption{
  Predicted (blue) and error-reduced experimental (orange) fidelities after removing the non-positive definite components using maximum likelihood estimation (MLE).
  The labels on the horizontal axis represent the error source to be reduced.
  The gray bar indicates the experimental fidelity before error reduction.}
\label{fig:param_contribution}
\end{figure}

Figure~\ref{fig:param_contribution} shows the expected fidelity with respect to the ideal density matrix if each error source is individually reduced from the experimental density matrix.
To achieve this, we calculate the expected density matrix with each error source reduced through the following two steps.
First, we compute the change in the density matrix due to each error source using $\Delta \rho_{\rm err} = \rho_{\rm sim} - \rho^*_{\rm sim}$, where $\rho^*_{\rm sim}$ is calculated based on the parameters listed in Table~\ref{tab:opt_val}, except that the parameter corresponding to the error source under consideration is set to its ideal-state value (see Sec.~\ref{sec:Optimizer}).
Second, we subtract the change obtained in step 1 from the experimental density matrix (i.e., $\rho_{\rm exp} - \Delta \rho_{\rm err}$).
We perform these steps for each error source to obtain the density matrices and simulate the reduction of the corresponding errors (see Appendix~\ref{sec:Maximum_likelihood_estimation} for the details of the MLE calculation of the simulated density matrices).
Figure~\ref{fig:param_contribution} shows that reducing the accidental coincidence error ($\eta$) produces the maximum improvement in fidelity.
Attenuating the pump to suppress multi-pair events is an effective countermeasure for the $\eta$ error, which is expected to improve fidelity by 4\%.
Besides that, the phase errors in the measurement bases ($\theta_{\rm AB}^{'+L} = \{\theta^{'+}_{\rm A},\ \theta^{'L}_{\rm A},\ \theta^{'+}_{\rm B},\ \theta^{'L}_{\rm B}\}$) make a large contribution.
The statistical fluctuations of the photon counting may have reduced the fitting accuracy of the coincidence curve, which determines the phase difference.
As a countermeasure, we calibrate at a pump intensity greater than that used for the calibration before the baseline experiment, thereby improving the fitting accuracy and reducing the error.
These measures are expected to improve fidelity by 2\%.

To examine the simulation results, we repeated the experiments that remedy the phase errors in the measurement bases and the accidental coincidence error (see Appendix~\ref{sec:Error-reduced_experiment}).
The experimentally obtained error-reduced density matrices are denoted $\rho_{\rm exp}^{\Diamond}$.
We calculate the fidelity of the experimentally obtained density matrix after error reduction followed by MLE, which we denote MLE($\rho_{\rm exp}^{\Diamond}$), with $\rho_{\rm ideal}$.
Figure~\ref{fig:param_contribution} shows that the obtained fidelity values are in good agreement with the predictions.
After reducing both of the phase errors in the measurement bases and the accidental coincidence error, the fidelity is 97\%, which is a 7\% improvement over the fidelity obtained before the error reduction (denoted by MLE($\rho_{\rm exp}$)).
This high-fidelity time-bin entangled state can be applied to specific quantum communication protocols, such as quantum key distribution~\cite{Honjo2008, Inagaki2013}, quantum teleportation~\cite{Takesue2015, Shen2023}, and entanglement swapping~\cite{Sun2017}.
If we assume that all the implemented errors are removed (see the value labeled ``all"  in Fig.~\ref{fig:param_contribution}), the expected fidelity is over 99\%, which means that the remaining ``After" trace distance shown in Fig.~\ref{fig:d_trace}(a) accounts for less than 1\% for the fidelity.

\section{Conclusion}
\label{sec:Conclusion}

We proposed the automated error-source quantification method, MBQEQ, and applied it to the QST experiment of the time-bin entangled photon-pair generation.
MBQEQ, which comprises a simulator, evaluator, and optimizer, allows us to estimate error sources in a human-understandable form from the experimental density matrix.
We modeled the relevant error sources using 26 model parameters and numerically simulated the reconstructed density matrix, including the modeled errors.
For the optimizer, we adopted the trace distance from the experimental density matrix as the cost function.
Starting from the ideal state, MBQEQ found the optimized model parameters that minimize the trace distance.
The reduction in the trace distance indicates that 86\% of the error sources were successfully attributed.
From the estimated parameters, we calculated the expected fidelity improvement when removing the errors and found that accidental coincidences and measurement-basis phase errors contributed significantly to the fidelity improvement.

Motivated by these results, we performed error-reduced experiments.
The obtained density matrices showed improved fidelities, consistent with the predictions.
Therefore, the proposed method enabled us to clarify the direction for optimizing the experimental system based on quantitatively estimated fidelity improvements, thereby contributing to the realization of high-quality quantum state generation. 
Further extensions of the method, such as implementing detector imperfections and external noises or applying MBQEQ to other experimental setups (e.g., quantum key distribution~\cite{Honjo2008, Inagaki2013}), are left for future work.
By adjusting the error model, the MBQEQ framework is applicable not only to quantum optics experiments but also to other platforms for quantum information experiments, such as superconducting qubits~\cite{Clarke2008, Devoret2013}, neutral atoms~\cite{Phillips1998, Henriet2020}, ion traps~\cite{Blatt2012, Bruzewicz2019}, and solid-state color centers~\cite{Awschalom2018}.
We also anticipate that this approach can be combined with advanced error-correction schemes or integrated into large-scale quantum network setups.

\begin{acknowledgments}
J.O. thanks Akihito Soeda and Yuki Sato for fruitful discussions on quantum information theory.
\end{acknowledgments}

\appendix
\section{Quantum state tomography}
\label{sec:Quantum_state_tomography}

We followed the derivation in Ref.~\cite{Takesue2009}.
Let $\{\Gamma_\mu\}$ denote the two-qubit Pauli matrices $\{I, \sigma_x, \sigma_y, \sigma_z\}^{\otimes 2}$ divided by 2.
Because Pauli matrices form the basis for density matrices, any density matrix for a two-qubit system can be written as a linear combination of Pauli matrices:
\begin{align}
\rho &= \sum_{\mu=1}^{16} \Gamma_\mu  {\rm Tr}\left(\Gamma_\mu \rho\right)
\equiv \sum_{\mu=1}^{16} \Gamma_\mu r_\mu.
\label{eq:rho_as_Gamma}
\end{align}
Because a two-qubit density matrix has 16 unknown parameters, we consider 16 measurements by the corresponding projection operators $\{\ket{\psi_\nu}\!\bra{\psi_\nu}\}$.
Defining $\ket{\psi_\nu} \in \{\ket{1}, \ket{2}, \ket{+}, \ket{L}\}^{\otimes 2}$, the measurement probability $s_\nu$ is 
\begin{align}
s_\nu = {\rm Tr}\left(\ket{\psi_\nu}\!\bra{\psi_\nu}\rho\right) = \bra{\psi_\nu}\rho\ket{\psi_\nu}.
\label{eq:s_nu}
\end{align}
In the experiment, $s_\nu$ is obtained from the coincidence counts in each projective measurement.
Note that $s_\nu$ is normalized so that the trace of the reconstructed density matrix is unity~\cite{James2001}.
Substituting the expression for $\rho$ into this gives
\begin{align}
s_\nu &= \sum_{\mu=1}^{16} \bra{\psi_\nu} \Gamma_\mu \ket{\psi_\nu} r_\mu 
\equiv \sum_{\mu=1}^{16} B_{\nu\mu} r_\mu.
\end{align}
The measurement bases $\{\ket{\psi_\nu}\}$ are chosen so that $B_{\nu\mu}$ becomes a nonsingular matrix, so the inverse matrix $B^{-1}$ exists. 
Therefore, we obtain the following expression for $r_\mu$:
\begin{align}
r_\mu = \sum_{\nu=1}^{16} (B^{-1})_{\mu\nu} s_\nu.
\end{align}
Substituting this back into Eq.~\eqref{eq:rho_as_Gamma} gives
\begin{align}
\rho &= \sum_{\mu=1}^{16} \Gamma_\mu \sum_{\nu=1}^{16} (B^{-1})_{\mu\nu} s_\nu
= \sum_{\nu=1}^{16} \left[\sum_{\mu=1}^{16} \Gamma_\mu (B^{-1})_{\mu\nu}\right] s_\nu
\equiv \sum_{\nu=1}^{16} M_\nu s_\nu.
\end{align}
Summarizing, the density matrix $\rho_{\rm QST}$ reconstructed by quantum state tomography can be given by
\begin{align}
\rho_{\rm QST} = \sum_{\nu=1}^{16} M_\nu s_\nu,
\end{align}
where $M_\nu$ is a matrix calculated as follows:
\begin{align}
M_\nu &= \sum_{\mu=1}^{16} \Gamma_\mu \,(B^{-1})_{\mu\nu},\\
B_{\nu\mu} &= \bra{\psi_\nu} \Gamma_\mu \ket{\psi_\nu}.
\end{align}
If we consider the measurement bases $\{\ket{\psi'_{\nu}}\}$ including phase and intensity errors, the measurement probabilities can change (i.e., $s'_\nu = \bra{\psi'_{\nu}}\rho\ket{\psi'_{\nu}} \neq s_\nu$). Meanwhile, the matrix $M_\nu$ is computed by assuming the ideal measurement bases $\{\ket{\psi_{\nu}}\}$, so the reconstructed density matrix $\rho'_{\rm QST} = \sum_\nu M_\nu s'_\nu$ is affected by errors in the measurement bases.
Even if the measurement bases are perfectly prepared, statistical fluctuations can still change the measurement probabilities (i.e., $s'_\nu = \bra{\psi_{\nu}}\rho\ket{\psi_{\nu}} + \delta_\nu \neq s_\nu$).

\section{Eigenvalue decomposition}
\label{sec:Eigenvalue_decomposition}

\begin{figure}[ht]
 \centering
 \includegraphics[width=0.99\linewidth]{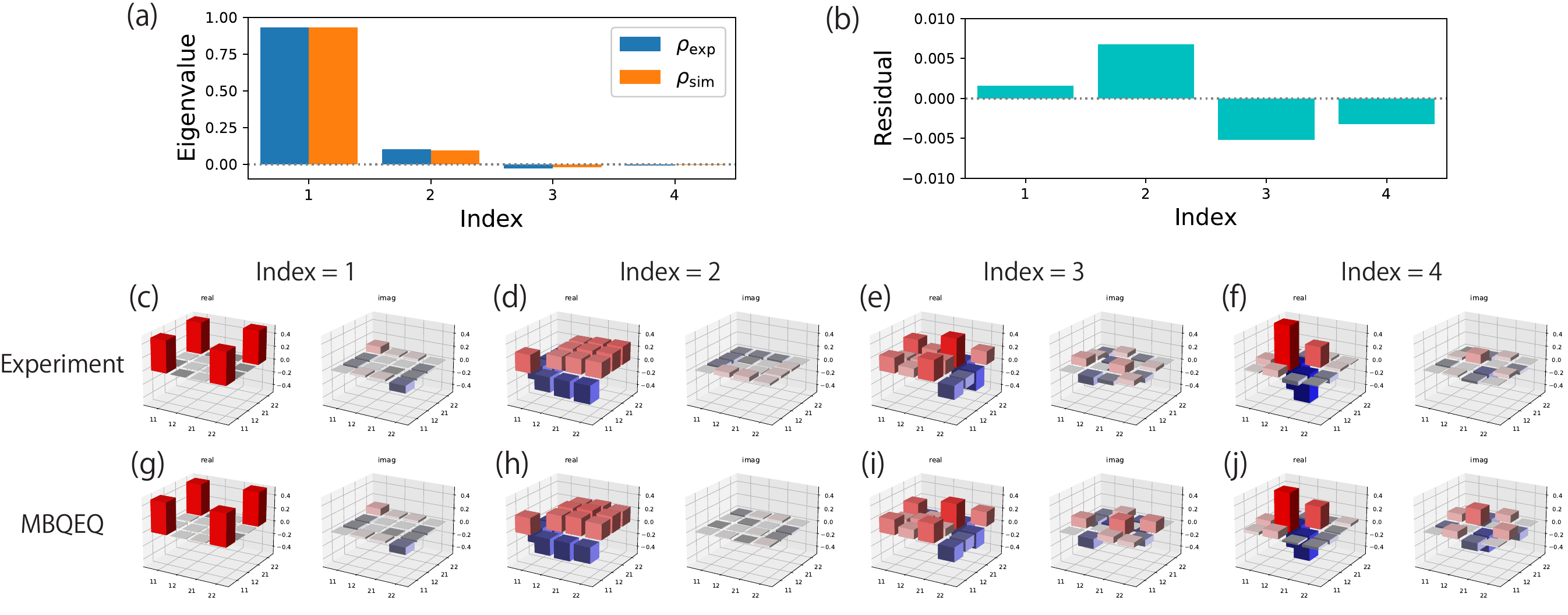}
  \caption{(a) Eigenvalues of experimental and simulated density matrices and (b) residuals. Eigenstates of (c)-(f) experimental and (g)-(j) simulated density matrices. The indices of the corresponding eigenvalues are shown above the panels.}
\label{fig:eig}
\end{figure}

Figure~\ref{fig:eig}(a) shows the eigenvalues of the experimental density matrix and the simulated density matrix obtained by MBQEQ.
Some eigenvalues are negative, which can arise from errors such as experimental imperfections or statistical fluctuations.
Figure~\ref{fig:eig}(b) shows the residuals for the two cases in Fig.~\ref{fig:eig}(a).
The differences are at most $\sim 1\%$.
In addition, Figs.~\ref{fig:eig}(c)-(j) compare the eigenstates, indicating that the two cases are in good agreement.

\section{Details of accidental coincidence error}
\label{sec:Details_of_accidental_coincidence_error}

To verify that the simulation results are consistent with the experimental results, we estimated the mean photon-pair number $\mu$ and the model parameter $\eta$ from the experimental data.
First, under our experimental conditions, we estimated the mean photon-pair number $\mu$ for each measurement, following the derivation in Ref.~\cite{Takesue2010}. 
Assuming that the number of photon pairs follows a Poisson distribution,
\begin{align}
P(\mu,x) = e^{-\mu}\frac{\mu^x}{x!},
\end{align}
the maximum and minimum coincidence count rates of the time-bin experiment can be expressed as follows:
\begin{align}
R_{\rm max} &\simeq \frac{1}{8}\mu\alpha_{\rm A}\alpha_{\rm B}
+ \left( \frac{1}{4}\mu\alpha_{\rm A} + d_{\rm A} \right)
\left( \frac{1}{4}\mu\alpha_{\rm B} + d_{\rm B} \right), \\
R_{\rm min} &\simeq \left( \frac{1}{4}\mu\alpha_{\rm A} + d_{\rm A} \right)
\left( \frac{1}{4}\mu\alpha_{\rm B} + d_{\rm B} \right).
\end{align}
Here, $\alpha_{\rm A(B)}$ and $d_{\rm A(B)}$ denote the detection efficiency and dark count rate per pulse, respectively.
The analytical expression for the visibility $V$ is
\begin{align}
V = \frac{R_{\rm max} - R_{\rm min}}{R_{\rm max} + R_{\rm min}}.
\end{align}
By substituting the experimental values for $\{\alpha_{\rm A},\ \alpha_{\rm B}\} = \{0.60,\ 0.27\}$ and $\{d_{\rm A},\ d_{\rm B}\} = \{2\times10^{-9},\ 2\times10^{-9}\}$, the visibility $V$ can be expressed as a function of the single variable $\mu$.
On the other hand, the expected visibility $V'$ can be calculated from the coincidence counts obtained in the QST experiment, which are proportional to $R_{\rm max}$ and $R_{\rm min}$.
Specifically, using the mean coincidence counts $\bar{n}_{\rm max}$ ($\propto R_{\rm max}$) of the projective measurements onto the states $\{\ket{11}_{\rm AB},\ \ket{22}_{\rm AB}\}$ and $\bar{n}_{\rm min}$ ($\propto R_{\rm min}$) of those onto the states $\{\ket{12}_{\rm AB},\ \ket{21}_{\rm AB}\}$, the experimental visibility can be estimated as,
\begin{align}
V' \simeq \frac{\bar{n}_{\rm max} - \bar{n}_{\rm min}}{\bar{n}_{\rm max} + \bar{n}_{\rm min}}.
\end{align}
By optimizing $\mu$ so that $V = V'$, we obtained the mean photon-pair number that reproduces the experimental coincidence counts.
Next, we estimated the model parameter $\eta$ from the experimental result.
Assuming that the generated state is a Werner state with $\epsilon$ as a parameter, that is,
\begin{align}
\epsilon \ket{\Phi}_{\rm AB}\!\bra{\Phi} + (1-\epsilon)\frac{I_2}{4},
\end{align}
and noting that the visibility is related to $\epsilon$ via $V' = \epsilon$,
the model parameter $\eta_{\rm exp}$ estimated from the experiment is
\begin{align}
\eta_{\rm exp} \simeq 1 - \epsilon = 1 - V'.
\end{align}

\begin{figure}[ht]
 \centering
 \includegraphics[width=0.5\linewidth]{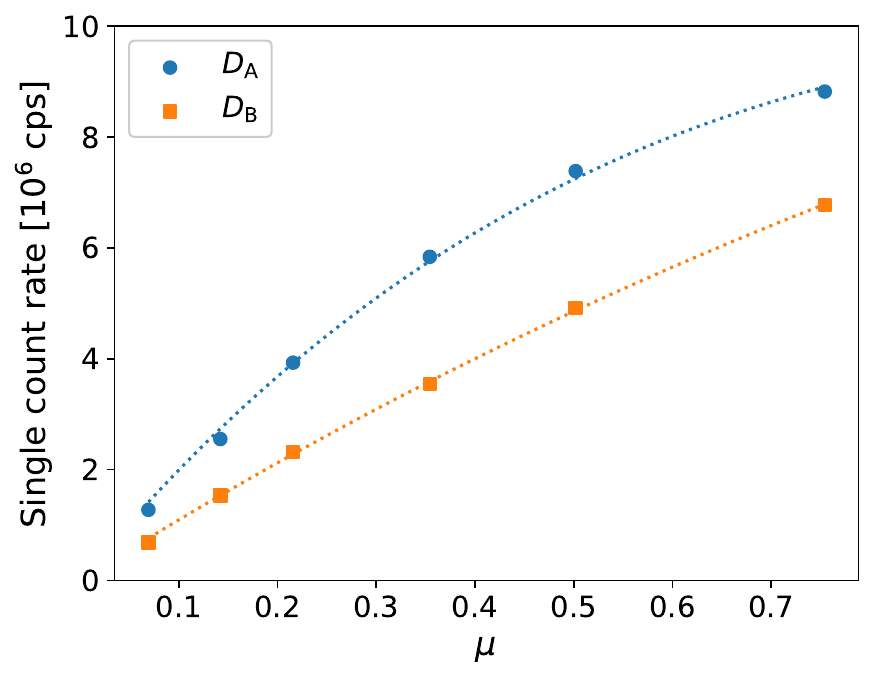}
  \caption{
  Single count rates for the detectors $D_{\rm A}$ and $D_{\rm B}$ as functions of the mean photon-pair number.
  Data points are experimental values. The dashed lines show fitting curves.}
\label{fig:single_count}
\end{figure}

If the accidental coincidence error is indeed due to the generation of multiple photon pairs, then the single count rate should be given by $\mu \xi f \exp(-\mu \xi f t_d/2)$~\cite{Takesue2005b}, where $f$ is the repetition frequency and $t_d$ is the dead time of the TIA.
By substituting the experimental values for $f = 500$ MHz and $t_d = 80$ ns,
and varying $\xi$ as the fitting parameter, we calculate fitting curves.
The single count rates of the two detectors ($D_{\rm A}, D_{\rm B}$) are shown in Fig.~\ref{fig:single_count} as a function of the mean photon-pair number, confirming that they are in good agreement with the fitting curves and justifying the model of accidental coincidence error in our experiments.

\section{Probability distribution of two-photon state}
\label{sec:Probability_distribution_of_two-photon_state}

\begin{figure}[ht]
 \centering
 \includegraphics[width=0.9\linewidth]{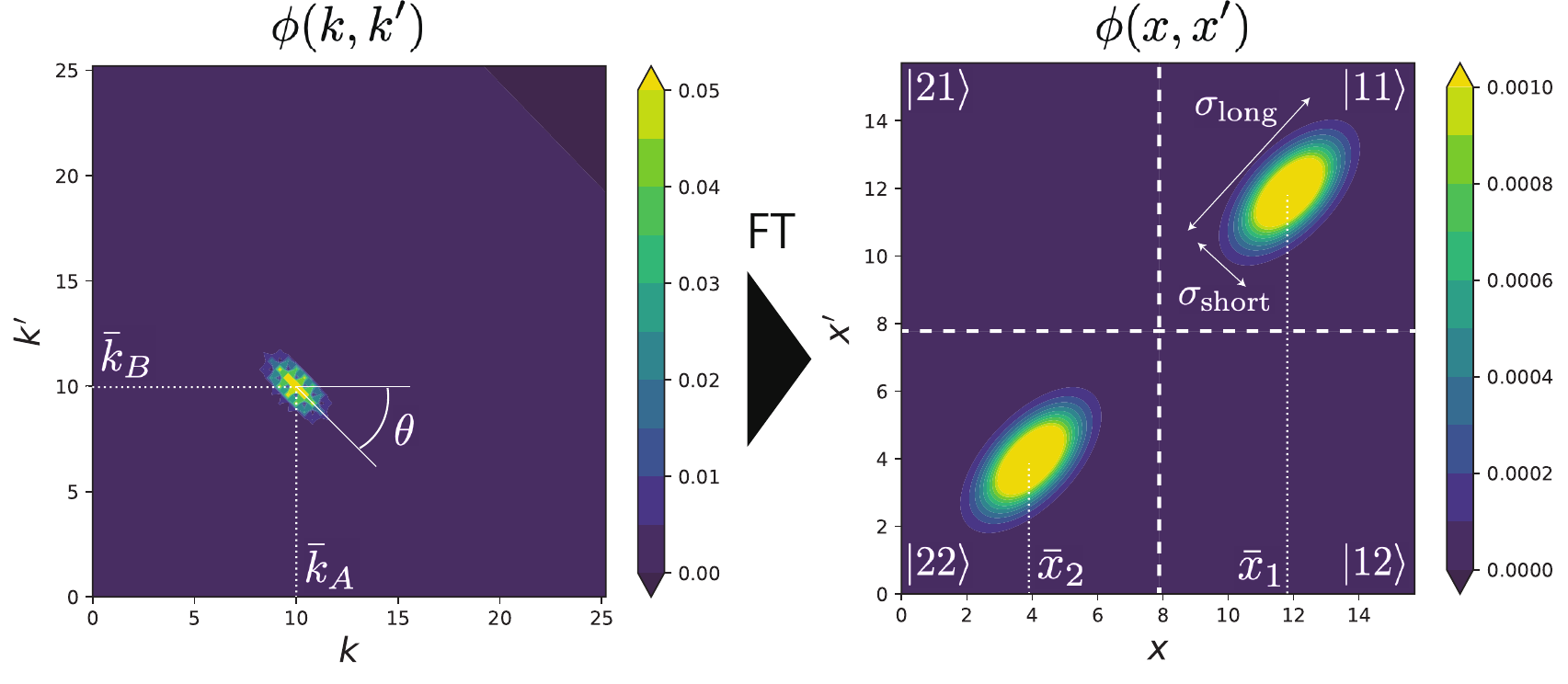}
  \caption{
  Probability distribution of two-photon state prepared in simulation.
  The left panel shows the distribution in $k$-space, and the right panel shows the distribution in real space obtained by Fourier transformation (FT).
  Parameters are $r_{\rm corr} = \sigma_{\rm long}/\sigma_{\rm short} = 2$, $p=0.5$, and $\theta = -\pi/4$.
  The contour shows the probability distribution.}
\label{fig:2ph_state}
\end{figure}

To analyze the probability distribution of the correlated photon pair, we used a quantum optics simulator~\cite{Oba2024}, a technique that enables the visualization of quantum phenomena involving photons and facilitates virtual experiments.
The probability distribution of the two-photon state in $k$-space is modeled by a Gaussian function as follows:
\begin{align}
\phi(k,k')
&= \sum_{i,j \in \{1,2\}} \alpha_{ij} f(k, k') \exp\left[- i (k \bar{x}_i + k' \bar{x}_j)\right]. 
\label{eq:gauss}
\end{align}
Here,
$\alpha_{11} = \sqrt{p}$,
$\alpha_{22} = e^{i \theta_{22}} \sqrt{(1-p)}$,
and $\alpha_{12} = \alpha_{21} = 0$
are the factor of each time-bin component in Eq.~\eqref{eq:Phi'_all}.
$\bar{x}_{1(2)}$ is the central position of the early (late) time bin in the simulation space.
The function $f(k,k')$ that determines the Gaussian shape is given by
\begin{align}
f(k,k') &= C \exp\left[-\frac{\sigma_{\rm short}^2 ((k - \bar{k}_A)\cos{\theta} + (k' - \bar{k}_B)\sin{\theta})^2}{2} \right] \nm \\
&\quad \times \exp\left[-\frac{\sigma_{\rm long}^2 (-(k - \bar{k}_A)\sin{\theta} + (k' - \bar{k}_B)\cos{\theta})^2}{2} \right]. 
\end{align}
Here, $\bar{k}_A$ and $\bar{k}_B$ are the central wave numbers of the two photons.
$\sigma_{\rm short}$ and $\sigma_{\rm long}$ are the pulse widths of the photon toward short and long axes, respectively.
$C$ is a normalization factor.
In this expression, the probability distribution of the two-photon state in $k$-space is rotated by $\theta$.
To impose the bosonic commutation relation, $\phi(k,k')$ is symmetrized by calculating $\phi(k,k') + \phi(k',k)$ with appropriate normalization.
By applying the Fourier transformation (FT) to the symmetrized $\phi(k,k')$, we obtain the real-space two-photon distribution $\phi(x, x')$, as shown in Fig.~\ref{fig:2ph_state}.
$\phi (x,x')$ is normalized as $\sum_{x,x'} |\phi (x,x')|^2 = 1$.
The interference pattern observed in the $k$-space distribution (see left panel of Fig.~\ref{fig:2ph_state}) reflects the fact that the real-space distribution is a superposition of the distribution with distinct central positions.

Including the information of $\phi (x,x')$ into the state $\ket{\Phi'}$ allows us to rewrite the state as
\begin{align}
\ket{\Phi'}_{\rm AB} &= \ket{\phi_{11}}\ket{11}_{\rm AB} + \ket{\phi_{12}}\ket{12}_{\rm AB}
+ \ket{\phi_{21}}\ket{21}_{\rm AB} + \ket{\phi_{22}}\ket{22}_{\rm AB},
\end{align}
where
\begin{align}
\ket{\phi_{11}} &= \sum_{n,n' = 0}^{N/2-1} \phi (x_n + L/2, x_{n'} + L/2) \ket{n}\ket{n'}, \\
\ket{\phi_{12}} &= \sum_{n,n' = 0}^{N/2-1} \phi (x_n + L/2, x_{n'}) \ket{n}\ket{n'}, \\
\ket{\phi_{21}} &= \sum_{n,n' = 0}^{N/2-1} \phi (x_n, x_{n'} + L/2) \ket{n}\ket{n'}, \\
\ket{\phi_{22}} &= \sum_{n,n' = 0}^{N/2-1} \phi (x_n, x_{n'}) \ket{n}\ket{n'}.
\end{align}
Here, $N$ and $L$ are the number of grids and the system length of the one-dimensional simulation space, respectively.
$\ket{n}$ represents a state in which a photon exists in the $n$th grid of each time bin.
In each time bin (i.e., the region separated by white-dashed line in Fig.~\ref{fig:2ph_state}), we define the lower-left corner as $(n,n') = (0,0)$ and the upper-right corner as $(n,n') = (N/2-1,N/2-1)$.
We also define position $x_n$ as $x_0 = 0$, $x_1 = L/N$, $\dots$, $x_{N/2} = L/2$.
For this representation of the two-photon state, the measurement probability $s_\nu$ in Eq.~\eqref{eq:s_nu} is 
\begin{align}
s_\nu &= {\rm Tr} \left[ \left(I \otimes P_\nu \right) \rho \right],
\end{align}
where $P_\nu = \ket{\psi_\nu}\!\bra{\psi_\nu}$, and the identity operator $I$ acts on the Hilbert space of the state $\ket{\phi_{ij}}$.

To reproduce the relative relationship between the experimental time-bin interval and pulse width, we set $N = 128$, $L = 5\pi$, $\bar{k}_A = 10 $, $\bar{k}_B =  (1547.1/1555.1) \bar{k}_A$, $\bar{x}_1 = 3L/4$, $\bar{x}_2 = L/4$, $\sigma_{\rm short} = (50/500)|\bar{x}_1-\bar{x}_2|$, and $\sigma_{\rm long} = r_{\rm corr} \sigma_{\rm short}$.
Here, $r_{\rm corr}$ is the model parameter representing the photon-pair correlation introduced in Sec.~\ref{sec:Correlation_of_photon_pair}.
In addition, we set $\theta = -\pi/4$ so that the real-space distribution tilts by $\pi/4$.

\section{Stability analysis}
\label{sec:Stability_analysis}

\begin{figure}[ht]
 \centering
 \includegraphics[width=0.9\linewidth]{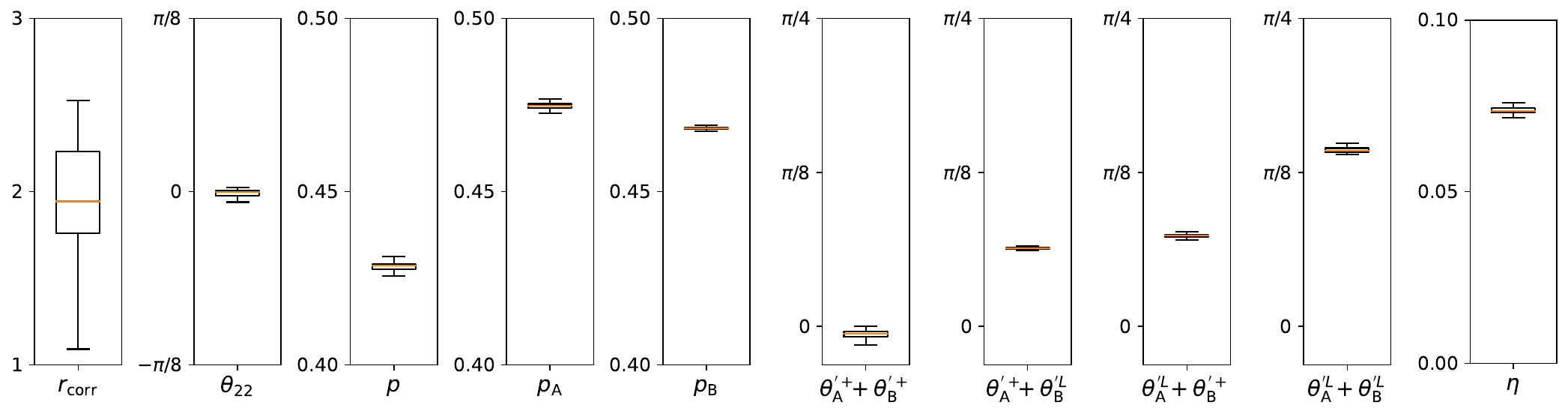}
  \caption{
  Box-and-whisker plots of the model parameters obtained from 100 independent optimizations with randomized initial conditions.
  The red line represents the median, the box encompasses the interquartile range (Q1-Q3), and the whiskers extend to the full range.
  The narrow boxes and short whiskers indicate high stability with respect to the initial condition.
  The statistical fluctuation parameters are omitted for brevity.}
\label{fig:stability}
\end{figure}

As described in Sec.~\ref{sec:Optimizer}, we analyzed the stability of the proposed method by performing MBQEQ 100 times with different initial guesses.
The target experimental density matrix was fixed by that of Fig.~\ref{fig:rho}(b).
Figure~\ref{fig:stability} shows box-and-whisker plots of the model parameters, showing that all parameters are robust against the choice of initial conditions, except for $r_{\rm corr}$.
As Fig.~\ref{fig:param_contribution} illustrates, $r_{\rm corr}$ contributes little to the density matrix, so a broad range of its values produces nearly identical density matrices, which reduces the robustness.

\section{Maximum likelihood estimation}
\label{sec:Maximum_likelihood_estimation}

We followed the derivation in Ref.~\cite{James2001}.
The physical candidate density matrix can be expressed as
\begin{equation}
  \rho_{\rm phys}(t)=\frac{T^{\dagger}(t)\,T(t)}
                    {\operatorname{Tr}\!\bigl[T^{\dagger}(t)T(t)\bigr]},
  \qquad
  t=(t_{1},\dots,t_{16})\in\mathbb{R}^{16},
\end{equation}
where $T(t)$ is the lower triangular matrix and is defined as follows:
\begin{align}
T(t)=
\begin{pmatrix}
 t_{1}               & 0                 & 0                 & 0 \\
 t_{5}+i\,t_{6}      & t_{2}             & 0                 & 0 \\
 t_{11}+i\,t_{12}    & t_{7}+i\,t_{8}    & t_{3}             & 0 \\
 t_{15}+i\,t_{16}    & t_{13}+i\,t_{14}  & t_{9}+i\,t_{10}   & t_{4}
\end{pmatrix}.
\end{align}
It guarantees trace normalization, Hermiticity, and positivity by construction.
Given a coincidence count data $\{n_{\nu}\}_{\nu=1}^{16}$,
the likelihood function to be minimized is defined as follows:
\begin{equation}
  \mathcal{L}(t)=
    \sum_{\nu=1}^{16}
      \frac{\bigl[\mathcal{N}\,p_{\nu}(t)-n_{\nu}\bigr]^{2}}
           {2\,\mathcal{N}\,p_{\nu}(t)},
   \label{eq:count_likelihood}
\end{equation}
where $p_{\nu}(t)=\bra{\psi_{\nu}}\rho_{\rm phys}(t)\ket{\psi_{\nu}}$ and $\mathcal{N}$ is a constant that depends on the experimental conditions such as photon flux and detector efficiency.

In numerical simulations, one typically starts from a (possibly non-positive definite) simulated density matrix $\rho_{\rm sim}$ without having access to $n_{\nu}$.
Therefore, we rewrite Eq.~\eqref{eq:count_likelihood} for the simulated density matrix by using $s_{\nu}=n_{\nu}/\mathcal{N}$, dividing the numerator and denominator of Eq.~\eqref{eq:count_likelihood} by $\mathcal{N}^{2}$, and eliminating the constant factor, which yields
\begin{equation}
  \ell(t)=
    \sum_{\nu=1}^{16}
      \frac{\bigl[p_{\nu}(t)-s_{\nu}\bigr]^{2}}
           {2\,p_{\nu}(t)}.
  \label{eq:prob_likelihood}
\end{equation}
Here, $s_{\nu}$ is the measurement probability defined by Eq.~\eqref{eq:s_nu} and is calculable for the simulation case.
Hence, we use Eq.~\ref{eq:prob_likelihood} as the likelihood function for the simulated density matrices (i.e., $\rho_{\rm exp} - \Delta \rho_{\rm err}$ discussed in Sec.~\ref{sec:Identification_of_high-impact_error_sources_by_ablation_study}).

We use Powell's method~\cite{Powell1964} implemented in the Python library \texttt{SciPy} (https://scipy.org) to obtain the optimized parameter $t^{\star}$.
For the initial guess, we use the same procedure given in Ref.~\cite{James2001}.
The resulting state
$\rho_{\mathrm{MLE}}=\rho_{\rm phys}(t^{\star})$ obeys
$\operatorname{Tr}\rho=1$, $\rho^{\dagger}=\rho$, and
$\rho\ge 0$.

\section{Error-reduced experiment}
\label{sec:Error-reduced_experiment}

We mainly address the phase error and accidental coincidence error.
For the phase error, to reduce the calibration error, we calibrated the system using a setup of which count rate was 2.5 times higher than that used for the baseline-experiment calibration, achieved by intensifying the pump pulses.
On the other hand, to reduce the accidental coincidence error, we performed the QST with a count rate 100 times lower than that of the baseline experiment by reducing the pump power, thereby decreasing the multi-pair generation rate.
The justification for modeling the accidental coincidence error in this manner is provided in Appendix~\ref{sec:Details_of_accidental_coincidence_error}.

\begin{figure}[ht]
 \centering
 \includegraphics[width=0.9\linewidth]{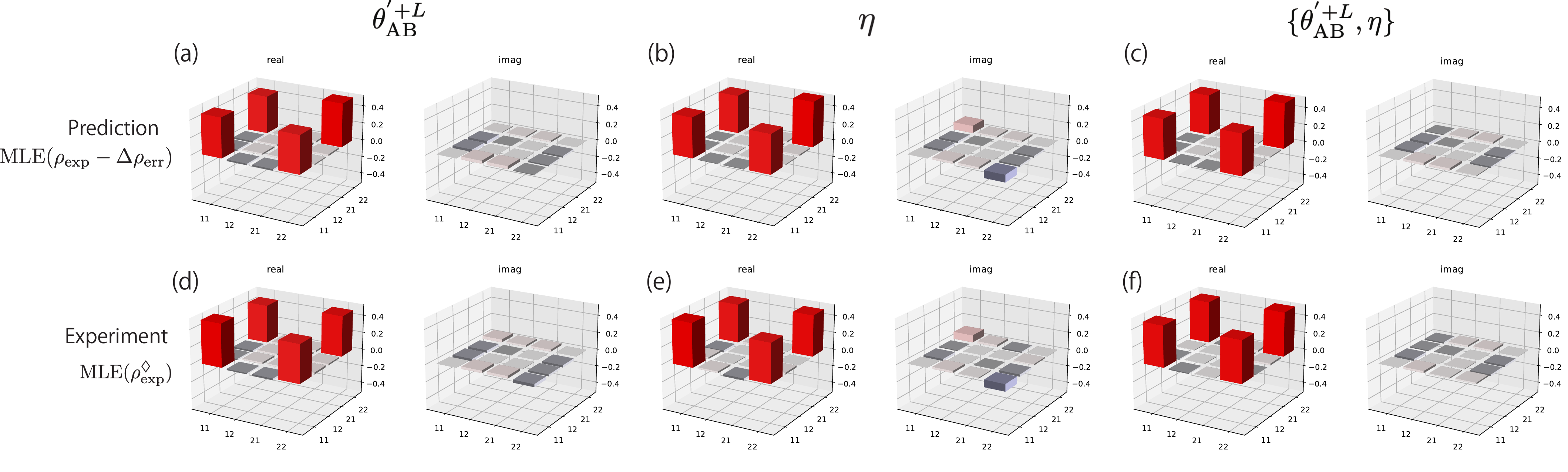}
  \caption{
  (a)-(c) Expected density matrices when the error source is removed.
  (d)-(f) Density matrices obtained from the error-reduced experiments.
  The error sources are indicated above the panels. For a description of each label, see Sec.~\ref{sec:Results_and_discussion}.
  }
\label{fig:rho_rm}
\end{figure}

Figure~\ref{fig:rho_rm} shows the density matrices predicted by our calculations and obtained experimentally after performing the error reduction.
Figure~\ref{fig:param_contribution} compares the resulting fidelities.

\end{document}